\documentclass[english,5p, number
]{elsarticle}
\usepackage[LGR,T1]{fontenc}
\usepackage{fancyhdr}
\usepackage{booktabs}
\usepackage{amssymb}
\usepackage{graphicx}
\usepackage{babel}

\makeatletter

\DeclareRobustCommand{\greektext}{%
  \fontencoding{LGR}\selectfont\def\encodingdefault{LGR}}
\DeclareRobustCommand{\textgreek}[1]{\leavevmode{\greektext #1}}
\ProvideTextCommand{\~}{LGR}[1]{\char126#1}

\providecommand{\tabularnewline}{\\}

\@ifundefined{date}{}{\date{}}
\usepackage{siunitx}
\usepackage{dblfloatfix}
\usepackage{capt-of}
\usepackage{cuted}
\usepackage{multicol}
\usepackage{balance}
\usepackage{url}

\makeatother

\begin{document}

\makeatletter
\def\ps@pprintTitle{% 
 \let\@oddhead\@empty 
 \let\@evenhead\@empty  
 \def\@oddfoot{}%  
 \let\@evenfoot\@oddfoot} 
\makeatother

\pagestyle{fancy}   
\fancyhf{}
\cfoot{\thepage}

\sloppy

\def\urlprefix{DOI }

\begin{frontmatter}

\title{Confined Continuous-Flow Plasma Source For High-Average-Power Laser
Plasma Acceleration}

\author{B. Farace, R. J. Shalloo, K. Põder, W. P. Leemans}

\address{Deutsches Elektronen-Synchrotron DESY, Notkestr. 85, 22607 Hamburg,
Germany \\
University of Hamburg, Department of Physics, Jungiusstr. 9, 20355
Hamburg, Germany }
\begin{abstract}
Over the last decades, significant advances in high-power laser systems
have enabled rapid progress in the development of laser-driven plasma
accelerators. Today, the results obtained in beam stability and reproducibility
present laser plasma acceleration as a viable and promising alternative
to conventional accelerators. As several electron beam and secondary
sources applications require high average currents, a major focus
is now on increasing the beam's repetition rate. In the following,
we introduce a novel plasma source for kHz electron acceleration,
providing a continuous and spatially confined gas flow, while minimising
the gas load in the acceleration chamber. 
\end{abstract}

\end{frontmatter}

Particle accelerators are a fundamental tool in science, medicine
and industry. In conventional accelerators, particles gain energy
in the electric fields generated between two electrodes or in an electromagnetic
(radio frequency) cavity, where accelerating fields as high as $\sim100$
MeV/m can be reached \citep{Shiltsev2021}. This value though, appears
to be the upper limitation: increasing the accelerating field further
will cause electrons to be stripped off from the cavity wall, leading
to what is usually addressed as ``electrical breakdown'' and damaging
the accelerating structure \citep{Grudiev2009}. Ever-increasing machine
sizes are then needed for generating higher energy particles. 

In the last decades, plasmas opened the path towards much higher electric
field gradients. In plasma-based accelerators an intense laser pulse,
or a high density particle beam, is propagated through a tenous plasma.
There, while the ions, in first approximation, can be considered at
rest, the free electrons, being less massive, are pushed away from
their original position by the ponderomotive (laser pulse) or electrostatic
(particle beam) force. They are then pulled back again by the restoring
force of the positive ions, yet their momentum makes them overshoot
and oscillate around their initial position. The pulse (or the beam),
hence, excites an electron oscillation inside the plasma, an electron
density wave, known as ``plasma wakefield'' \citep{Tajima1979},
as depicted in Fig. 1. As a fully ionized plasma
cannot suffer from electrical breakdown, this wave can sustain much
higher field gradients, up to hundreds of GeV/m \citep{Esarey2009},
promising to shrink the machine size by a factor 1000. High quality
electron beams in the 100 MeV range were already produced in plasma
accelerators nearly two decades ago \citep{Mangles_2004,Geddes_2004,Faure_2004}.
The GeV scale was reached soonly after \citep{Leemans_2006}, paving
the way towards the multi-GeV regime \citep{Kim_2013,Wang_2013,Leemans_2014}.
\begin{center} 
\begin{tabular}{c} 
\vspace{0.47cm}   
\includegraphics[width=8.5cm]{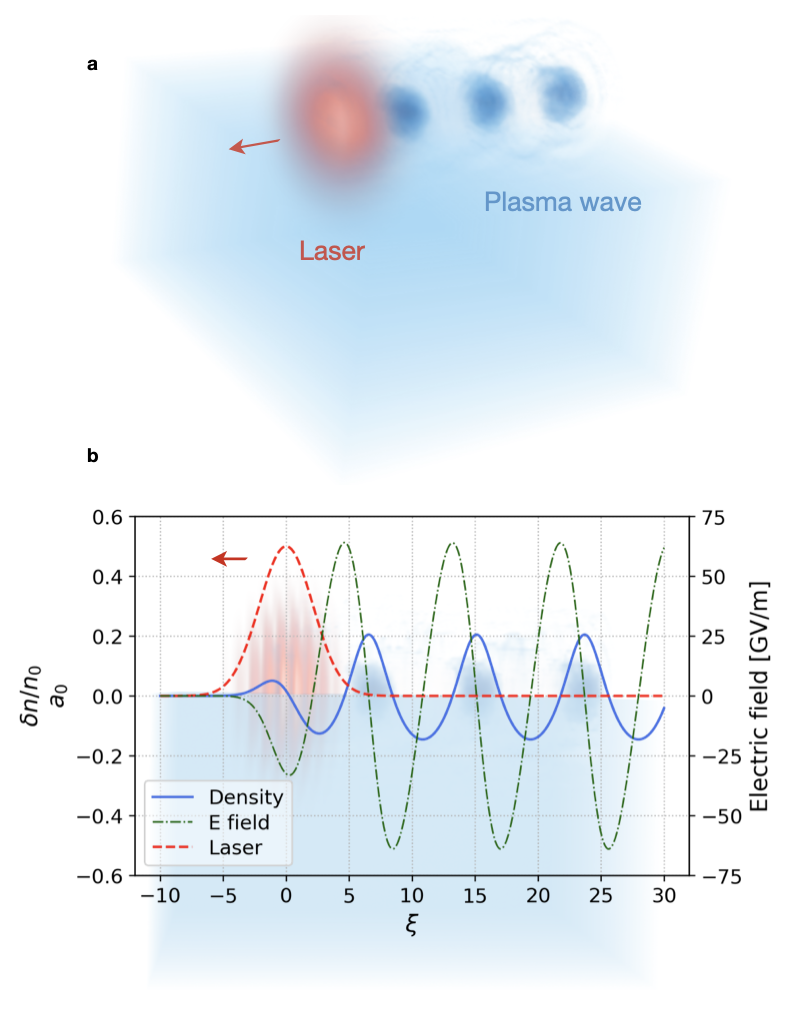}   
\end{tabular}   
\captionof{figure}{In panel (a), a laser pulse of moderate intensity (red) is made to propagate, from right to left, inside a plasma (blue) and excites a trailing linear plasma wave. The same interaction is depicted in panel (b) from a different angle,  showing the density perturbation $\delta n/n_0$ (blue) and the resulting electric field  (green) along the spatial coordinate $\xi$.} 
\label{wakefield}

\vspace{0.47cm}
\end{center}
\begin{table*}
\begin{centering}
\begin{tabular}{cccccccc}
\toprule 
{\footnotesize{}$E$} & {\footnotesize{}\textgreek{t}} & {\footnotesize{}$a_{0}$} & {\footnotesize{}$n_{e}$} & {\footnotesize{}$\Delta E$} & {\footnotesize{}$w_{0}$} & {\footnotesize{}$Z_{R}$} & {\footnotesize{}$L_{dep}$}\tabularnewline
\midrule
\midrule 
{\footnotesize{}\SI{30}{\joule}} & {\footnotesize{}\SI{60}{\femto\second}} & {\footnotesize{}$4.8$} & {\footnotesize{}\SI{0.66e18}{\per\centi\metre\cubed}} & {\footnotesize{}\SI{4.2}{\giga\electronvolt}} & {\footnotesize{}\SI{26}{\micro\metre}} & {\footnotesize{}\SI{2.6}{\milli\metre}} & {\footnotesize{}\SI{4.5}{\milli\metre}}\tabularnewline
\midrule 
{\footnotesize{}\SI{1}{\joule}} & {\footnotesize{}\SI{25}{\femto\second}} & {\footnotesize{}$3.5$} & {\footnotesize{}\SI{4.2e18}{\per\centi\metre\cubed}} & {\footnotesize{}\SI{500}{\mega\electronvolt}} & {\footnotesize{}\SI{10}{\micro\metre}} & {\footnotesize{}\SI{0.4}{\milli\metre}} & {\footnotesize{}\SI{2.8}{\milli\metre}}\tabularnewline
\midrule 
{\footnotesize{}\SI{3}{\milli\joule}} & {\footnotesize{}\SI{5}{\femto\second}} & {\footnotesize{}$2$} & {\footnotesize{}\SI{1e20}{\per\centi\metre\cubed}} & {\footnotesize{}\SI{10}{\mega\electronvolt}} & {\footnotesize{}\SI{2.1}{\micro\metre}} & {\footnotesize{}\SI{18}{\micro\metre}} & {\footnotesize{}\SI{25}{\micro\metre}}\tabularnewline
\bottomrule
\end{tabular}
\par\end{centering}
\caption{{\footnotesize{}\label{Tab. 1}Laser plasma accelerators scaling in
the blowout regime, assuming $\lambda_{0}=800$ nm, adapted from \citep{Faure_2018}.
The main parameters that define the laser-plasma interaction are outlined
for different energy regimes ($E$): the pulse temporal duration $\tau$,
the normalised vector potential $a_{0}$, the electron density $n_{e}$,
the achievable energy gain $\Delta E$, the laser spot size $\omega_{0}$,
the Rayleigh length $Z_{R}$, and the dephasing length $L_{dep}$.}}
\end{table*}
\begin{figure*}
[!b]
\begin{centering}
\includegraphics[scale=0.49]{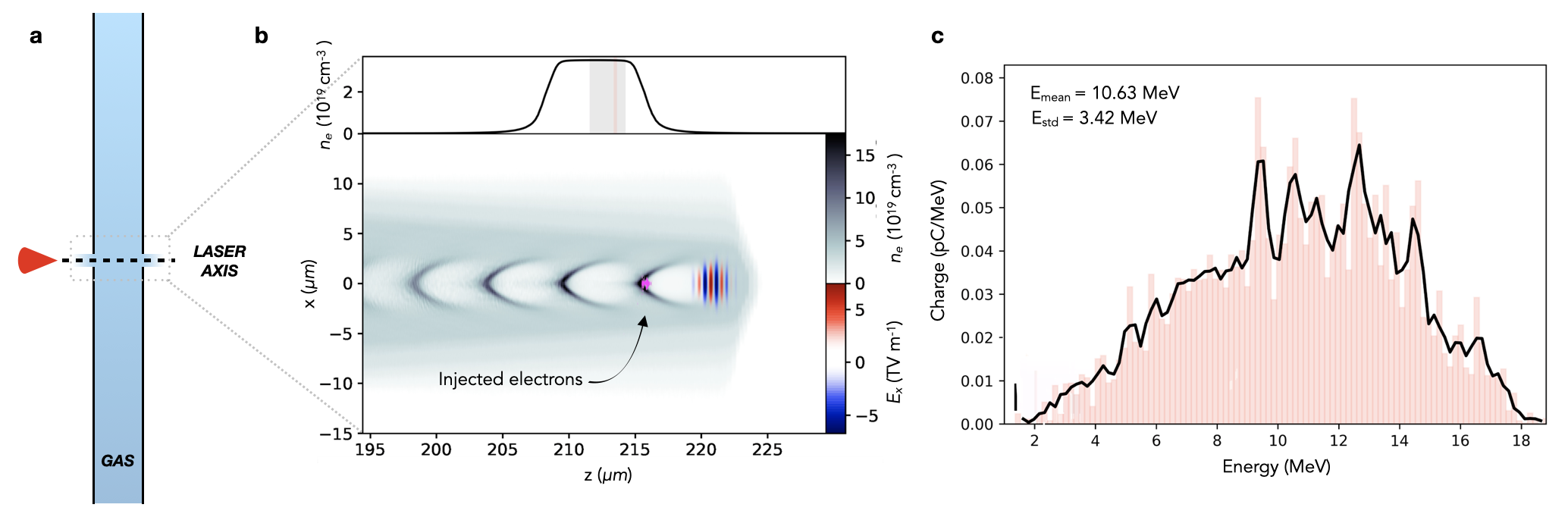}
\par\end{centering}
\caption{{\footnotesize{}\label{Fig. 2}Plasma source schematics. The laser
intersects the capillary transversely, drives a plasma wave and accelerates
electrons (a). Panel (b) shows the results of a PIC simulation. The
upper section depicts a typical density profile obtainable with the
presented source geometry (simulated in ANSYS-FLUENT). In the lower
one, a mJ laser pulse with the parameters of Tab. 1 is made to propagate
through it. Injected electrons are highlighted in pink. Panel (c)
shows the electrons spectrum at the end of the propagation. The total
accelerated charge in this example is around 0.5 pC. The PIC simulation
has been carried out with FBPIC \citep{Lehe2016}.}}
\end{figure*}
To date, electrons up to 8 GeV have been produced, employing a Petawatt-class
laser system focused into a 20 cm long plasma channel \citep{Gonsalves_2019}.

Among the different challanges that this technique is facing \citep{Wang2021,Lindstrom2021,Diederichs2020,Buscher2020},
increasing the repetition rate of the machines is essential, in order
to drive applications where many pulses per second are needed. 

In laser-driven plasma accelerators (LPAs), 1 kHz operation can now
be routinely achieved \citep{Faure_2018} and, on this path, great
effort is being made in optimizing the plasma source, which is key
in the acceleration process. The source geometry, as well as its gas
density and composition, have a major impact on the accelerated electron
beam properties, including charge, energy spectrum and divergence.
In the following, we present a novel plasma source which provides
a continuous, controllable, 100 micron scale confined gas region.
Its interaction with an intense laser pulse is able to produce few
MeV electron beams at kHz repetition rates.\medskip{}

Currently, most plasma-based accelerators are operated in the ``blow-out''
regime, where the fields generated in the plasma wave are particularly
suitable for electron acceleration \citep{Pukhov2002,Lu2006}. This
regime requires highly relativistic intensities, which are usually
delivered by focusing down joule-class Ti:Sa laser systems, typically
operating with repetition rates \ensuremath{\lesssim} 10 Hz. High
repetition rate plasma accelerators, on the other hand, make use of
kilohertz laser systems, where the maximum pulse energy is limited
to the millijoule level. With these lower energies the blow-out regime
is still accessible, provided that all the other parameters are correctly
scaled. As an example, Tab. 1 shows the characteristics of a typical
plasma source suitable for such systems. In order to accelerate electrons
at kHz repetition rates, the plasma source must provide high gas density
($n_{e}\sim$ \SI{1e20}{\per\centi\metre\cubed}) and must be limited
to tens of micrometers scale \citep{Gu_not_2017,Salehi2021,Rovige2020,Gustas2018}.
Such sources are usually gas nozzles, with inner diameter < \SI{100}{\micro\metre}.
Because of their bigger outer diameter, though, these are very prone
to be damaged by the tightly focused driving laser, therefore the
laser height above the nozzle must be finely tuned and is typically
$\sim$ \SI{100}{\micro\metre}. This inevitably means that a very
high backing pressure is needed (up to 60 bar \citep{Gustas2018})
in order to achieve the desired gas density. Gas load inside the acceleration
chamber is then one of the major issues is such experiments, potentially
limiting the applicability of these novel accelerators.\balance

The novel source proposed here is able to supply continuous gas flow
(required for kHz operation) in a confined spatial region at the \SI{100}{\micro\metre}
scale. It consists of a hollow, thin-walled, micro-capillary tube,
where gas is flowing at a controlled pressure. As shown in Fig. 2,
the driving laser intersects the capillary transversely, ionizing
the gas, exciting the plasma wave and providing electron injection.
The gas is in a closed system for capture and recirculation and, thus,
the leakage into the chamber can be minimised. Controlling the input
pressure allows, then, for tuning the desired gas density at the interaction
point and the transverse geometry results in very short ramps, which
are needed for optimising the coupling of the driver into the gas.
Moreover, features could be added inside the capillary in order to
induce shocks or other flow manipulations which could improve the
acceleration process. 

The novel plasma source proposed, hence, promises to increase the
LPA stability and optimise the gas load during the production of few
MeV electron beams at kHz repetition rate. Such systems have the potential
to deliver beams of femtoseconds duration with high average currents,
which would be extremely useful in several medical and industrial
applications \citep{Glinec2005,Sciaini2011}. \balance

\bibliographystyle{unsrtnat}
\bibliography{Plasmasource}

\end{document}